\documentclass[prb,twocolumn,draft,amsmath,showpacs]{revtex4}
\usepackage{graphics}\input epsf
\begin{document}

\title{Magnetic excitations and phonons in the spin-chain compound NaCu$_2$O$_2$}

\author{K.-Y. Choi,$^{1}$ V. P. Gnezdilov,$^2$
P. Lemmens,$^{3,4}$ L. Capogna,$^{5,6}$ \\
M. R. Johnson,$^6$ M.
Sofin,$^3$ A. Maljuk,$^3$ M. Jansen,$^3$ and B. Keimer$^3$}

\affiliation{$^1$ Institute for Materials Research, Tohoku
University, Katahira 2-1-1, Sendai 980-8577, Japan}

\affiliation{$^2$ B. I. Verkin Institute for Low Temperature
Physics NASU, 61164 Kharkov, Ukraine}

\affiliation{$^3$ Max Planck Institute for Solid State Research,
D-70569 Stuttgart, Germany}

\affiliation{$^4$ Institute for Physics of Condensed Matter, TU
Braunschweig, D-38106 Braunschweig, Germany}

\affiliation{$^5$ CNR-INFM, CRS-SOFT and OGG Grenoble, 6 Rue J.
Horowitz, BP 156 F-38042 Grenoble CEDEX 9, France}

\affiliation{$^6$ Institut Laue Langevin, 6 rue J. Horowitz  BP
156 38042 Cedex 9 Grenoble, France}

\date{\today}

\begin{abstract}
We report an inelastic light scattering study of
single-crystalline NaCu$_2$O$_2$, a spin-chain compound known to
exhibit a phase with helical magnetic order at low temperatures.
Phonon excitations were studied as a function of temperature and
light polarization, and the phonon frequencies are compared to the
results of {\it ab-initio} lattice dynamical calculations, which
are also reported here. The good agreement between the observed
and calculated modes allows an assignment of the phonon
eigenvectors. Two distinct high-energy two-magnon features as well
as a sharp low-energy one-magnon peak were also observed. These
features are discussed in terms of the magnon modes expected in a
helically ordered state. Their polarization dependence provides
evidence of substantial exchange interactions between two closely
spaced spin chains within a unit cell. At high temperatures, the
spectral features attributable to magnetic excitations are
replaced by a broad, quasielastic mode due to overdamped spin
excitations.
\end{abstract}

\maketitle

\section{Introduction}
A large research effort over the past three decades has been
devoted to the exploration of quantum effects in spin-chain
materials. At low temperatures, most quasi-1D magnets exhibit
either gapped spin-liquid ground states, or states with collinear
long-range magnetic order. \cite{Lemmens03} In some spin-chain
compounds, spin-lattice interactions stabilize spin-Peierls ground
states. Recently, several quasi-1D magnets with noncollinear,
incommensurate long-range order at low temperatures have also been
found. LiCu$_2$O$_2$, an insulating compound containing
edge-sharing CuO$_2$ chains, is a prominent example.
\cite{Zvyagin02,Masuda04,Gippius,Masuda05,choi,comments}
Initially, both lattice disorder \cite{Masuda04} and spin
frustration \cite{Gippius} had been discussed as the driving
mechanism of the observed helimagnetic order. This controversy now
appears to have been resolved:\cite{comments} The incommensurate
magnetic ordering is due to long-range competing exchange
interactions along the CuO$_2$ chains. Questions remain, however,
about the microscopic role of the substantial number of Li
impurities substituting copper ions on the spin chains, and about
the magnitude of the exchange interactions between two closely
spaced CuO$_2$ chains within a unit cell. Inelastic neutron
scattering measurements \cite{Masuda05} suggest that these
interactions are substantial, whereas electronic structure
calculations \cite{Gippius} indicate much smaller values.

Investigations of the disorder-free, isostructural compound
NaCu$_2$O$_2$ \cite{Maljuk04} have now begun to help resolve these
and other questions. \cite{Capogna05} This material is also
interesting because it is the end member of the series
Na$_{1+x}$Cu$_{2}$O$_{2}$. For nonzero $x$, the spin chains of
these materials are doped and exhibit intriguing phenomena such as
Wigner crystallization of holes. \cite{Horsch05} Like its
Li-analog, NaCu$_{2}$O$_{2}$ has an orthorhombic crystal structure
with space group Pnma. The lattice parameters are a=6.2087~\AA,
b=2.9343~\AA, and c=13.0548~\AA.~\cite{Maljuk04} Magnetic
Cu$^{2+}$ ions form two inequivalent edge-sharing copper oxide
chains per unit cell. They run along the {\it b} axis and are
shifted relative to each other by $b/2$. Along the $a$ and
$c$-directions, they are separated by nonmagnetic Na$^{+}$ and
Cu$^{+}$ ions, respectively. The Cu-O-Cu bond angle is
92.9$^{\circ}$. This results in a small, ferromagnetic,
nearest-neighbor (NN) exchange interaction. The antiferromagnetic
next nearest-neighbor (NNN) interaction is the largest parameter
in the spin Hamiltonian. An analysis of high-temperature
susceptibility demonstrates the significance of interactions
reaching up to fourth-nearest neighbors of a Cu$^{2+}$ ion.
\cite{Capogna05} The static susceptibility displays a broad
maximum around 52 K, associated with the evolution of short-range
antiferromagnetic correlations. \cite{Capogna05} Upon cooling,
susceptibility and specific heat measurements show two magnetic
phase transitions at 12 and 8 K. Neutron diffraction results
\cite{Capogna05} demonstrate that the former transition is due to
a noncollinear magnetic ordering with the propagation vector (0.5,
$\zeta$, 0) with $\zeta=0.227$, similar to that of LiCu$_2$O$_2$
where $\zeta=0.174$. \cite{Masuda04} This underscores the
conclusion that lattice disorder (which is negligible in
NaCu$_{2}$O$_{2}$) is not responsible for the helimagnetism in
this class of compounds. The ordered moment of 0.56 $\mu_B$ is
reduced with respect to the full spin-1/2 moment due to quantum
fluctuations. The latter transition might be related to spin
canting.

While both Raman scattering \cite{choi} and inelastic neutron
scattering \cite{Masuda05} work on magnetic excitations in
LiCu$_2$O$_2$ has been reported, the magnetic dynamics of
NaCu$_{2}$O$_{2}$ has thus far not been investigated. In this
article, we report an investigation of magnetic excitations in
NaCu$_{2}$O$_{2}$ by polarized Raman scattering. In the
magnetically long-range ordered state, three distinct features
attributable to magnetic excitations are observed. Two modes
around 18 and 58 cm$^{-1}$ are observed both with parallel
in-chain and crossed polarizations. This points to significant
interchain interactions. In addition, a broad maximum around 390
cm$^{-1}$ is present in in-chain polarization only. The
temperature dependences of the low-energy and the two high-energy
modes are characteristic of one- and two-magnon excitations,
respectively. Upon heating, the three magnetic modes disappear,
and we observe a pronounced quasielastic response predominantly in
the in-chain polarization channel. These results are discussed in
the light of previous work on CuO$_2$ spin chain materials,
especially recent neutron scattering data and model calculations
on LiCu$_2$O$_2$. \cite{Masuda05}

The phonon spectrum of NaCu$_{2}$O$_{2}$ was also investigated as
a function of temperature and light polarization. The observed
modes agree well with those predicted by {\it ab-initio} lattice
dynamical calculations, which allows us to identify the phonon
eigenvectors at the zone center. Unusually strong two-phonon modes
with an unconventional temperature dependence are also observed.
These features may be related to a broad electronic mode around
500 cm$^{-1}$, which possibly arises from electronically active
defects.

This article is organized as follows. In Section II, we describe
technical details pertaining to the Raman experiments and the
lattice dynamical calculations. Sections III.A and III.B contain
Raman data on phonon and magnetic excitations, respectively, along
with a discussion of possible mode assignments. Section IV
provides a summary of our findings.

\section{Experimental details and lattice dynamical calculations}

The single crystals  were grown using a self-flux method as
described elsewhere. \cite{Maljuk04} The chemical composition of
the studied sample is close to the ideal stoichiometry, as
discussed previously. \cite{Capogna05} Raman scattering
measurements were performed in quasi-backscattering geometry with
the excitation line $\lambda= 514.5$ nm of an Ar$^{+}$ laser.  The
laser power of 7 mW was focused to a 0.1 mm diameter spot on the
sample surface. Stokes and anti-Stokes measurements confirmed that
the incident radiation did not increase the temperature of the
samples by more than a few K. Spectra of the scattered radiation
were collected by a DILOR-XY triple spectrometer and recorded by a
nitrogen cooled charge-coupled device detector with a spectral
resolution less than $\sim 1$ cm$^{-1}$.

The optical phonon frequencies were calculated within the density
functional theory using the VASP software package. \cite{Kresse}
Projector augmented wave (PAW) potentials with an energy cutoff of
283, 259, and 273 eV were used for O, Na, and Cu, respectively.
The atomic coordinates were first optimized in the single unit
cell. Then a (2,4,1) supercell, which is 2, 4, and 1 times larger
in the respective a, b and c directions, was constructed and the
coordinates of the 160 atoms were optimized again. For the
dispersion relation calculations, two k-points in the $a$ and $b$
directions and one in the $c$ direction were retained, while for
the (2,4,1) supercell a single k-point was used. Supercell
calculations are required in order for the magnitude of the force
constants to become negligible (which happens beyond ~10
Angstroms) as well as to calculate the phonon dispersions exactly
instead of using an interpolation. A series of single point energy
calculations, which give the Hellmann-Feynmann forces acting on
all atoms in the supercell, were then performed on the structure
obtained by displacing, one at a time, each of the 20 inequivalent
atoms in positive and negative directions along the Cartesian
axes. Finally, the dynamical matrix for any point in reciprocal
space was generated and diagonalised.

\section{Results and discussion}
\subsection{Phonons}

Figure~1 displays Raman spectra in $bb$-, $ba$-, and $aa$-
polarizations  at 5 and 295 K. Most of the modes shown are of
lattice vibrational origin. In order to assign these modes, we
performed a group theoretical analysis as well as {\it ab-initio}
lattice dynamical calculations following the description in
Section II. NaCu$_{2}$O$_{2}$ has the space group Pnma, where all
atoms have a 4c site symmetry. Apart from the acoustic (B$_{1u}$ +
B$_{2u}$ + B$_{3u}$) modes, the factor group analysis yields the
following Raman- and infrared-active modes: 10 A$_g$(aa,bb,cc) + 5
B$_{1g}$(ab) + 10 B$_{2g}$(ac) + 5 B$_{3g}$(bc) + 5 A$_{u}$ + 9
B$_{1u}$ + 4 B$_{2u}$ + 9 B$_{3u}$. Thus, in $bb$- and
$aa$-geometries we expect 10 A$_g$ modes, and in the $ba$-
configuration 5 B$_{1g}$ modes.

The observed peak frequencies are listed in Table I together with
the calculated ones. Given that the calculation was performed
without adjustable parameters, the agreement of observed and
calculated frequencies is quite satisfactory. We can thus
tentatively assign the observed modes to specific vibration
patterns. Figures 2 and 3 provide a synopsis of the eigenvectors
of the Raman-active modes resulting from the calculation. It can
be seen that the highest-energy modes around 500 cm$^{-1}$ are
predominantly oxygen vibrations, as expected based on the small
mass of the oxygen ion.

%%%%%%%%%%%%%%%%%%%%%%%%%%%%%%%%%%%%%%%%%%%%%%%%%%%%%%%%%%%%%%%%%
%%%%%%%%%%%%%%%%%TABLEI%%%%%%%%%%%%%%%%%%%%%%%%%%%%%%%%%%%%%%%%%%
\begin{table}
\caption{Comparison between experimental and calculated phonon
frequencies of the Raman-active modes in a unit of cm$^{-1}$.}
\label{TablI}
  \centering
\begin{tabular}{cc|cccc}
\hline \hline \multicolumn{2}{c|}{observed frequencies} &
\multicolumn{4}{c}{calculated frequencies}
\\
\,(aa),(bb)\, & \,\,\,\,(ab)\,\,\,\,\,\,\,\, &
\,\,\,\,\,\,10A$_g$\,\,\,\, & \,\,\,\,5B$_{1g}$\,\,\,\,
& \,\,\,\,10B$_{2g}$\,\,\,\, & \,\,\,\,5B$_{3g}$\,\,\,\,\\
\hline

55.7    & 55.7    & 42.2    & 41.4   & 47.2   &    \\

        &         &         &        & 80.6   & 79  \\

114.6   &  114.6  &    93.5 &        &   91.8 &  109   \\

142.5   & 142.5   &  131.7  &        &        &    \\

146.2   & 146.2   &  149.6  &  147.23 &  161.5 &    \\

164.6   & 164.6   & 181.1   &  169.3 &  197.3 &  167.9 \\

217.5   &         &  254.5  &  253.1 & 253.4  & 289 \\

317.4   &  317.4  &  405.8  &      &    393    &      \\

432.9   &         &  439.7    &  425.7   & 433.5  &  413  \\

467.3   & 451.7   &         &        &  473   &      \\

501.9   & 501.9   &  510.2  &        &        &      \\

556.8   & 556.8   &  551.9  &        &        &   \\

566.6   & 566.6   &         &        &        &      \\
571.6   & 571.6   &         &        &   585.9  &      \\
\hline \hline
\end{tabular}
\end{table}
%%%%%%%%%%%%%%%%%%%%%%%%%%%%%%%%%%%%%%%%%%%%%%%%%%%%%%%%%%%%%%%%%%%%
%%%%%%%%%%%%%%%%%%%%%%%%%TABLEI%%%%%%%%%%%%%%%%%%%%%%%%%%%%%%%%%%%%%%

In both $aa$- and $bb$-polarizations we observe a total of 13
A$_g$ modes. This number is larger than the expected 10 modes.
Specifically, the highest-energy oxygen vibration around 560
cm$^{-1}$ seems to be split into three modes, indicating different
environments of oxygen ions in the crystal structure (see Fig. 7).
This is either due to coherent deviations from the reported Pnma
lattice symmetry too subtle to be discernible in the reported
x-ray and neutron \cite{Capogna05} powder diffraction data, or to
lattice defects. In $ab$-polarization 11 modes are observed, a
number that is also larger than the symmetry-allowed 5 B$_{1g}$
modes. Our calculations show that the estimated frequency
separation between the A$_g$ and B$_{1g}$ modes is only
$1-2\,\mbox{cm}^{-1}$. This lies within our spectral resolution of
$\sim 1$ cm$^{-1}$. It is therefore difficult to assess whether
the modes observed in $ab$-polarization are pure B$_{1g}$ modes,
or whether they are due to an admixture of A$_g$ modes due to
polarization leakage caused by lattice defects and/or other
mechanism.

The temperature dependence of the peak frequencies and linewidths
of representative phonons, extracted by fitting the corresponding
spectra to Lorentzian profiles, is shown in Fig. 4. Nearly all
phonon modes exhibit a similar behavior. With decreasing
temperature the peak frequencies harden and the linewidths narrow
gradually, before saturating at about 70 K. This behavior is
commonly observed as a consequence of lattice anharmonicity.
However, two phonons exhibit phenomena not attributable to
anharmonic effects alone. First, the linewidth of the 146
cm$^{-1}$ mode, which according to the calculation mostly involves
a $b$-axis vibration of the Cu$^{1+}$ ions against the rest of the
lattice (Fig. 3), exhibits a subtle but distinct drop upon cooling
from 50 K to the lowest temperatures (see the arrow in Fig. 4). As
the CuO$_2$ chains vibrate in unison, a strong spin-phonon
coupling is not expected, and the relationship of this subtle
anomaly to the development of magnetic correlations is not
obvious. Second, the 502 cm$^{-1}$ oxygen vibration shows an
asymmetric lineshape, which appears to be due to an anti-resonance
with a broad electronic mode centered around 500 cm$^{-1}$. The
origin of the electronic mode will be addressed in Section III.B
below.

Here we draw attention to the unusual behavior of the two-phonon
peak around 1000 cm$^{-1}$, which is probably an overtone of the
502 cm$^{-1}$ mode (Fig. 1). In $bb$-polarization and at 5 K, the
integrated intensity of the two-phonon peak is nearly an order of
magnitude higher than that of the one-phonon peak. At room
temperature, the two-phonon peak {\it decreases} strongly in
intensity. In contrast, the ratio of the integrated intensities of
one- and two-phonon modes for the $aa$-configuration is only $\sim
0.3$ in the temperature regime investigated. Since two-phonon
excitations are usually weak and hardly change with increasing
temperature, the behavior of this mode is quite surprising.
Qualitatively similar (albeit less pronounced) effects were
observed in corner-sharing copper oxide chain systems, and
interpreted in terms of a Fr\"{o}hlich interaction mechanism.
\cite{misochko,abrashev} This mechanism can also lift the
selection rules such that infrared-active modes may become
observable by Raman spectroscopy. This might explain the observed
extra phonon modes. A quantitative understanding of the two-phonon
modes of NaCu$_{2}$O$_{2}$ remains a subject of future work.

\subsection{Magnetic excitations}

Magnetic excitations in the Raman spectra can be identified by
virtue of their temperature and polarization dependences. The
upper panel of Fig. 1 shows a substantial temperature-induced
rearrangement of spectral weight in in-chain $bb$-polarization.
The low-energy part of the spectrum that is affected by this
redistribution is highlighted in Fig. 5. Upon cooling, a broad
quasielastic background is progressively replaced by a more
structured spectrum. At 5 K, three features attributable to
magnetic excitations emerge: (i) an $\sim 50$ cm$^{-1}$ wide peak
centered at 390 cm$^{-1}$; (ii) a weaker peak with an asymmetric
lineshape extending from about 50 to 170 cm$^{-1}$; (iii) and a
sharp peak around 17 cm$^{-1}$. The latter two features are also
present in $ab$-polarization (see the middle panel of Fig. 1).
Their temperature dependence is highlighted in Figs. 6 and 7,
respectively. The magnetic Raman operator is given by $\mathcal{R}
\propto S_i \cdot S_j$, \cite{Lemmens03} so that magnetic Raman
scattering is dominant in the polarization where there are
substantial exchange paths. The observation of features (ii) and
(iii) in both $bb$- and $ba$-polarizations thus indicates that
they are appreciably affected by interchain coupling.

Before discussing these features in detail, we recall the salient
properties of the spin Hamiltonian of NaCu$_{2}$O$_{2}$.
\cite{Capogna05} The largest exchange coupling parameter is the
antiferromagnetic NNN coupling along the spin chains ($J_2 = 62.6$
cm$^{-1}$). The helix structure is induced by a frustrating
ferromagnetic NN interaction $J_1 = 11.4$ cm$^{-1}$. Longer-range
exchange interactions along the chains and/or interactions between
spins within a closely spaced pair of spin chains are also
required to obtain a good fit to the magnetic susceptibility and
the propagation vector of the helix structure.

The magnetic excitations of NaCu$_{2}$O$_{2}$ have thus far not
been studied, but an inelastic neutron scattering study has
recently addressed the low-energy spin excitations in
LiCu$_{2}$O$_{2}$. \cite{Masuda05} Based on model calculations,
\cite{Mizuno} the largest parameter in the spin Hamiltonian,
$J_2$, is expected to be very similar in the two systems. From the
$\sim$30\% smaller pitch angle of the magnetic helix in
NaCu$_{2}$O$_{2}$ one can infer a correspondingly smaller NN
coupling $J_1$. This is expected, because the Cu-O-Cu bond angle
of NaCu$_{2}$O$_{2}$ is closer to 90$^\circ$, and it is consistent
with the data currently available on this compound. Although there
may be further subtle differences between the spin Hamiltonians of
both compounds, the neutron scattering data and model calculations
reported for LiCu$_{2}$O$_{2}$ can thus serve as a useful
guideline for the interpretation of our data. \cite{Masuda05} Note
that the Raman data presented here cover the high-energy and
low-energy segments of the spin wave spectrum not probed by the
neutron experiments on LiCu$_{2}$O$_{2}$. Both data sets can thus
be regarded as complementary in this respect as well.

We first discuss the two broad modes centered at 70 and 390
cm$^{-1}$. The temperature dependence of the former feature is
shown in more detail in Fig. 5. It persists well into the
paramagnetic state and merges into the background at temperatures
exceeding about 80 K. The 390 cm$^{-1}$ feature persists to even
higher temperatures in excess of 150 K, without a strong
renormalization of its lineshape (Fig. 7). This is the temperature
dependence expected for two-magnon Raman scattering in
low-dimensional magnets, where short-range spin correlations
persist well into the paramagnetic state. The presence of the
higher-energy two-magnon peak in $bb$-polarization, and its
absence in $ab$-polarization, confirm that the magnetic
zone-boundary energy is determined predominantly by exchange
interactions within a single spin chain, as expected based on the
discussion above. Its energy is expected to be slightly below
twice the magnetic zone-boundary energy, due to magnon-magnon
interactions. This is indeed the case, as the zone-boundary energy
computed in Ref. \onlinecite{Masuda05} is 26 meV ($= 210$
cm$^{-1}$).

The presence of a second two-magnon feature is more unusual, but
it can be understood as a consequence of the helical spin
correlations present in NaCu$_{2}$O$_{2}$ and LiCu$_{2}$O$_{2}$.
The helical spin structure and the four-atom basis in the magnetic
unit cell give rise to a complex spin wave spectrum (Fig. 6a of
Ref. \onlinecite{Masuda05}), with several magnon branches
exhibiting either maxima or minima in the energy range 8-12 meV
(65-97 cm$^{-1}$) for LiCu$_{2}$O$_{2}$. Because of the lower
value of $J_1$ (see discussion above), this energy scale is
expected to be reduced in NaCu$_{2}$O$_{2}$. The high two-magnon
density of states due to the vanishing spin-wave velocity at these
extrema of the dispersion relation naturally explains the second
two-magnon Raman peak. The observation of this peak in both $bb$-
and $ab$-polarization indicates that magnons in the energy range
around 10 meV are significantly influenced by interchain exchange
interactions within a pair of closely spaced chains. This agrees
with the findings of Ref. \onlinecite{Masuda05}, but seems
incompatible with the prediction of negligible interchain exchange
interactions based on LDA calculations. \cite{Gippius}

We will now address the origin of the peak at 18 cm$^{-1}$. As
shown in detail in Fig. 6, the peak broadens and shifts to lower
frequency upon heating, and it vanishes above the magnetic
transition temperature. This behavior is characteristic of
one-magnon excitations. A one-magnon excitation with almost
identical energy has been observed in CuGeO$_3$, a spin-Peierls
compound also based on edge-sharing copper oxide chains, in
magnetic fields high enough to stabilize a phase with an
incommensurate modulation of both the spin density and the crystal
lattice. \cite{Loa} However, the presence of both spin and lattice
modulations has made it difficult to arrive at a conclusive
interpretation of the nature of this mode and the origin of its
Raman-activity. \cite{Loa,Enderle,Uhrig} One-magnon modes of
comparable energies have also been observed in Raman scattering
experiments on the two-dimensional antiferromagnet La$_2$CuO$_4$.
\cite{Gozar}  These zone-center transverse magnon modes are Raman
active by virtue of spin-orbit coupling, which manifests itself in
an exchange anisotropy two orders of magnitude smaller than the
isotropic Cu-O-Cu superexchange interaction.

The most straightforward interpretation of the 18 cm$^{-1}$ peak
is based on a transverse magnon at $q=0$. This is consistent with
neutron scattering measurements on LiCu$_{2}$O$_{2}$, which
tentatively indicate a magnon gap of comparable magnitude.
\cite{Masuda05} If this scenario is valid, the size of this gap
constitutes a large fraction of the isotropic superexchange
interaction. This may at first seem surprising in view of the
nearly quenched orbital moment of the Cu$^{2+}$ ion. However,
prior work on edge-sharing copper oxide chain systems has
experimentally demonstrated substantial superexchange
anisotropies. \cite{Kataev} The surprisingly large relative
magnitude of the non-isotropic part of the exchange interaction
can be understood based on electronic structure calculations for
the 90$^\circ$ Cu-O-Cu bond geometry. \cite{Kataev} A zone-center
transverse magnon is therefore a plausible origin of the 18
cm$^{-1}$ peak.

An alternative interpretation of this peak is in terms of a
longitudinal magnon. Such excitations are usually heavily damped,
but can live long enough to be probed near quantum critical
points. Indeed, longitudinal magnons have recently been observed
in Raman scattering experiments \cite{Gros} on the coupled
spin-tetrahedra system Cu$_2$Te$_2$O$_5$Br$_2$, which exhibits
incommensurate helical order and a spin excitation spectrum not
unlike those of NaCu$_{2}$O$_{2}$. \cite{Zaharko05,Peter01}
Because of the sizable ordered moment of NaCu$_{2}$O$_{2}$,
\cite{Capogna05} this scenario is less plausible. However, a
conclusive assignment of the 18 cm$^{-1}$ peak will have to await
neutron spectroscopy experiments on NaCu$_{2}$O$_{2}$ akin to
those that have led to the identification of longitudinal magnons
in other systems. \cite{Tun,Zheludev,Lake} Another open question
concerns the temperature dependence of the frequency and intensity
of the 18 cm$^{-1}$ mode. In contrast to one-magnon excitations in
most collinear magnets, whose frequency and intensity roughly
scale with the order parameter, we observe a temperature-linear
behavior of these quantities (Fig. 6). Detailed model calculations
for the observed helical magnetic structure are required to
address the origin of this behavior.

At high temperatures, the one- and two-magnon peaks are replaced
by an intense quasielastic peak due to overdamped spin
excitations. Similar phenomena have been studied in other
quasi-one-dimensional magnets such as CuGeO$_3$.
\cite{Kuroe97,Brenig} Using a hydrodynamic description of the
correlation function, \cite{Lemmens03} the quasielastic Raman peak
is well described by a Lorentzian profile $I(\omega)\propto
C_{m}T^{2}D_{T}k^{2}/[{\omega^{2} +(D_{T}k^{2})^{2}}]$, where $k$
is the scattering wave vector, $D_{T}$ the thermal diffusion
constant, and $C_{m}$ the magnetic specific heat. Figure 8
displays a representative fit of the quasielastic response to a
Lorentzian profile at 295 K. The inset of Fig. 8 shows the
temperature dependence of the scattering amplitude, which
increases rapidly with increasing temperature and then saturates
around 200 K. This is the same temperature at which the
higher-energy two-magnon peak vanishes, heralding the
disappearance of short-range spin-spin correlations. Here we note
that deviations from the Lorentzian profile are discernible above
100 cm$^{-1}$ (not shown here). While this is partly due to an
ambiguity in determining the scattering background, it may also
reflect interference from an additional scattering channel at
higher energies, which we will now discuss.

Figure 7 indeed shows a broad, weakly temperature independent peak
centered around 500 cm$^{-1}$, the tail of which may well extend
down to about 100 cm$^{-1}$. The peak is observed almost
exclusively in the $bb$-polarization channel, although some traces
may also be present in the other polarizations. As already
mentioned in Section III.A, the oxygen vibration at 502 cm$^{-1}$
exhibits a pronounced anti-resonance behavior with this mode. The
peak does not shift as the laser frequency is varied, so that
luminescence can be ruled out. Since the strong anisotropy, the
nearly temperature dependent spectral weight, and the featureless
shape of the peak are inconsistent with multiphonon scattering, it
is probably of electronic origin.

An unambiguous assignment of this peak is not possible based on
the data at hand. A priori, one has to consider orbital, charge,
and magnetic excitations of the electrons. Raman scattering
experiments have recently uncovered evidence of orbital
excitations in transition metal oxides with $t_{2g}$ valence
electrons at energies ranging from $\sim 500$ to $\sim 2000$
cm$^{-1}$, \cite{miyasaka,ulrich} but orbital excitations for the
$e_{g}$ electron of a Cu$^{2+}$ ion are expected to occur at much
higher energies. Because of the presence of both Cu$^{1+}$ and
Cu$^{2+}$ ions in NaCu$_2$O$_2$, one might also consider
low-energy charge excitations as the origin of the peak. The
eigenvectors of both of the phonons exhibiting anomalous behavior
suggest that the corresponding vibrations modulate
Cu$^{1+}$-Cu$^{2+}$ charge-transfer excitations. However, band
structure calculations and photoemission data for LiCu$_2$O$_2$
indicate considerably larger energies for such excitations.
\cite{Zatsepin} Finally, a spinon continuum extending beyond the
magnon bandwidth is expected on general grounds in one-dimensional
magnets. Indeed, Raman scattering from such continuum excitations
has been reported for edge-sharing copper oxide chains.
\cite{misochko} In this scenario, the phonon anomalies discussed
in Section III.A may be consequences of a spin-phonon interaction,
also invoked to explain phonon anomalies in CuGeO$_3$.
\cite{Loa96} While we consider this as the most likely
interpretation of the 500 cm$^{-1}$ peak, we cannot rule out
entirely that it is a consequence of defects in the NaCu$_2$O$_2$
crystal structure. As the Na/Cu ratio determined in prior work on
NaCu$_2$O$_2$ was stoichiometric within an error margin of about
2\%, \cite{Capogna05} defects on these atomic sites can be present
only in very small concentrations. Oxygen defects can, however,
not be ruled out. Whatever their origin, it is conceivable that
electronically active lattice defects could give rise to charge
carriers in shallow traps, which shake off spin excitations
propagating along the $b$-direction when excited into the
conduction band, giving rise to a broad Raman peak. Further work
with other probes is required to conclusively establish the origin
of the 500 cm$^{-1}$ peak.

\section{Conclusions}

To conclude, we have presented a detailed study of the phonon
spectrum and the magnetic response of the helically ordered
spin-chain system NaCu$_2$O$_2$ using Raman spectroscopy. Based on
lattice dynamical calculations, most of the phonon modes could be
assigned. Several spectral features originating from magnetic
excitations were observed as well. The rich magnetic Raman
spectrum of NaCu$_2$O$_2$ is unusual for a quasi-one-dimensional
magnet. However, a comparison to model calculations and neutron
scattering data on the isostructural compound LiCu$_2$O$_2$ and
prior Raman scattering work on other copper oxide chain systems
yields plausible interpretations of all of the observed modes.

\section*{Acknowledgments}
We thank Manuel Cardona, Peter Horsch, and Karin Schmalzl for
useful discussions, and D. Richard and A. Filhol for technical
assistance.

\clearpage

%%%%%%%%%%%%%%%%%%%%%%%%%%%%%%%%%%%%%%%%%%%%%%%%%%%%%%%%%%%%%
%Figure 1
%%%%%%%%%%%%%%%%%%%%%%%%%%%%%%%%%%%%%%%%%%%%%%%%%%%%%%%%%%%%%
\begin{figure}[t]
\begin{center}
\leavevmode \epsfxsize=8cm \epsfbox{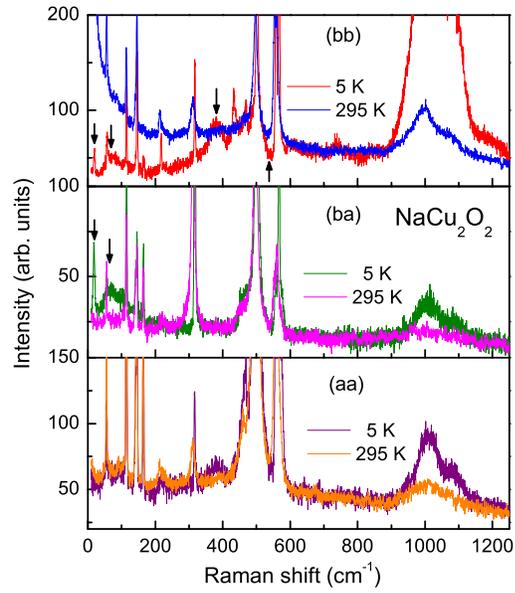} \caption{(Color
online) Polarization dependence of Raman spectra at 5 and 295 K in
$bb$-, $ba$-, and $aa$- polarizations, respectively. The downward
vertical arrows indicate magnetic excitations while the upward one
denotes a Fano-shaped dip corresponding to the phonon at 502
cm$^{-1}$. } \label{fig.1}
\end{center}
\end{figure}
%%%%%%%%%%%%%%%%%%%%%%%%%%%%%%%%%%%%%%%%%%%%%%%%%%%%%%%%%%%%%%%

%%%%%%%%%%%%%%%%%%%%%%%%%%%%%%%%%%%%%%%%%%%%%%%%%%%%%%%%%%%%%
%Figure2
%%%%%%%%%%%%%%%%%%%%%%%%%%%%%%%%%%%%%%%%%%%%%%%%%%%%%%%%%%%%%
\begin{figure}[t]
      \begin{center}
       \leavevmode
       \epsfxsize=7cm
       \epsfbox{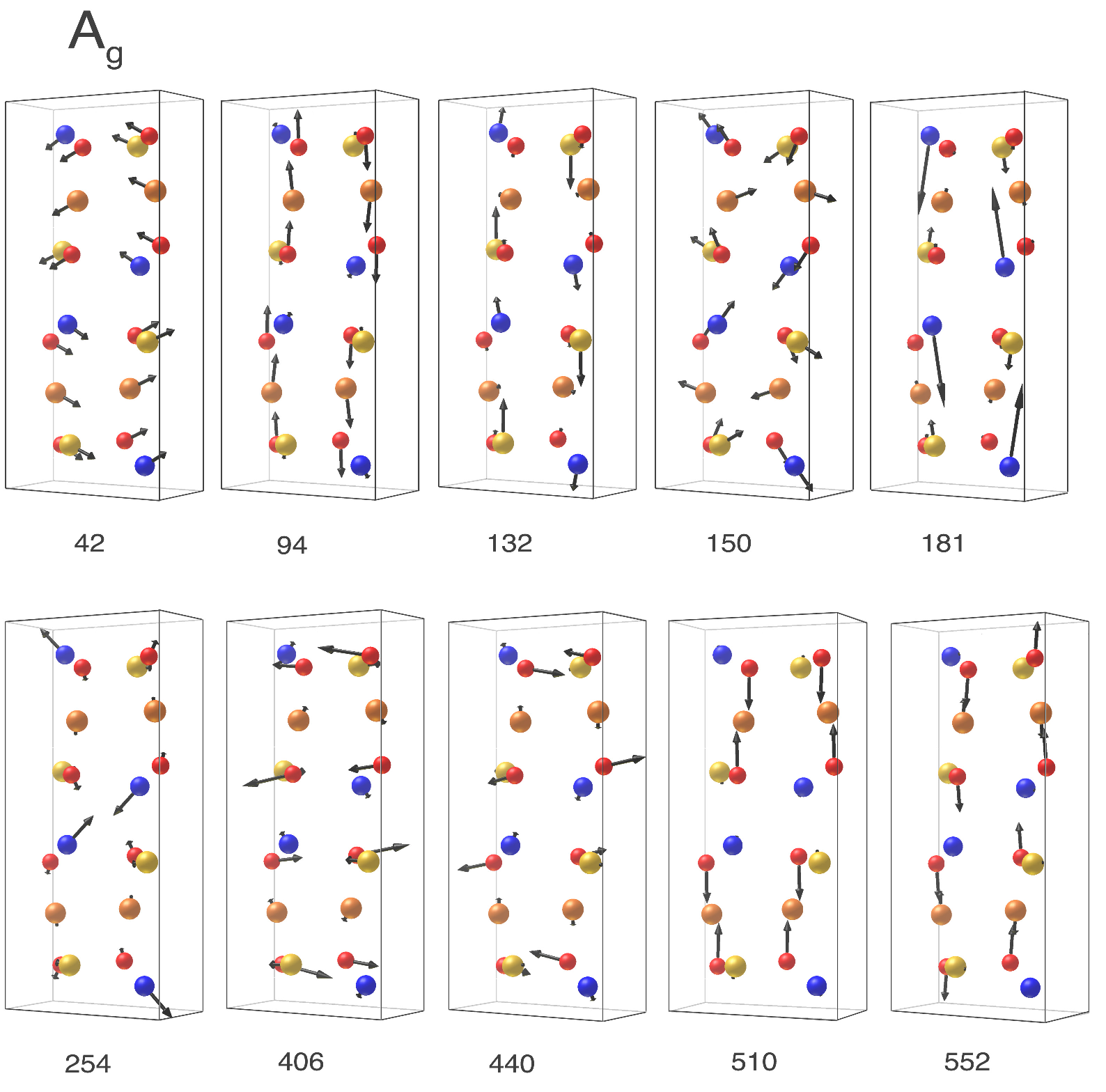}
        \caption{(Color online) Calculated normal frequencies and eigenvectors
of the $A_g$-symmetry modes.  The relative amplitude of the
vibrations (arrow length) is correct within each mode. For the
sake of clarity a multiplication factor has been applied to some
modes. The red balls stand for the oxygen ions, the blue ones for
sodium, the yellow ones for the Cu$^{+2}$, and the orange ones for
the Cu$^{+1}$.} \label{fig.2}
\end{center}
\end{figure}
%%%%%%%%%%%%%%%%%%%%%%%%%%%%%%%%%%%%%%%%%%%%%%%%%%%%%%%%%%%%%%%%%

%%%%%%%%%%%%%%%%%%%%%%%%%%%%%%%%%%%%%%%%%%%%%%%%%%%%%%%%%%%%%
%Figure3
%%%%%%%%%%%%%%%%%%%%%%%%%%%%%%%%%%%%%%%%%%%%%%%%%%%%%%%%%%%%%
\begin{figure}[t]
      \begin{center}
       \leavevmode
       \epsfxsize=7cm
       \epsfbox{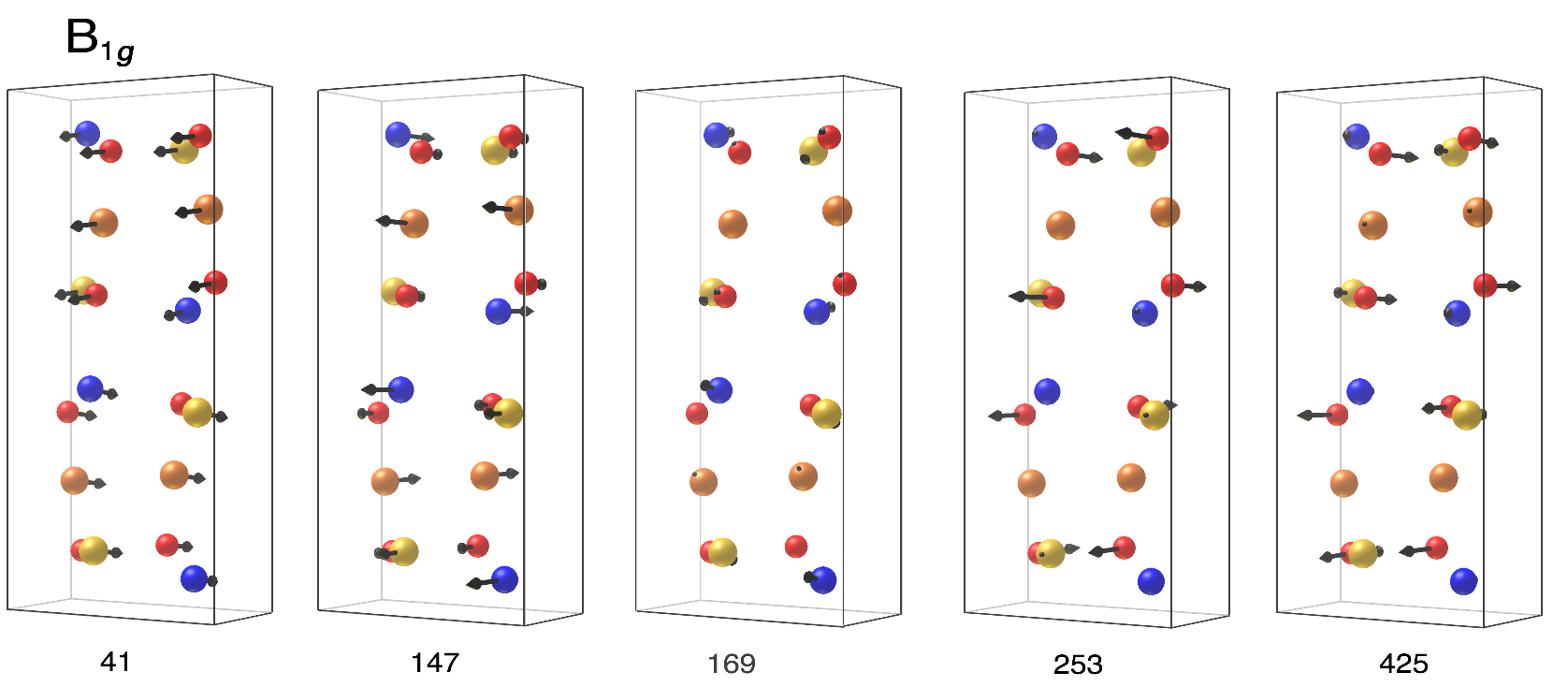}
        \caption{(Color online) Calculated normal frequencies and eigenvectors
 of the $B_{1g}$-symmetry modes. The relative amplitude of the vibrations
(arrow length) is correct within each mode. For the sake of
clarity a multiplication factor has been applied to some modes.}
\label{fig.3}
\end{center}
\end{figure}
%%%%%%%%%%%%%%%%%%%%%%%%%%%%%%%%%%%%%%%%%%%%%%%%%%%%%%%%%%%%%%%%%

%%%%%%%%%%%%%%%%%%%%%%%%%%%%%%%%%%%%%%%%%%%%%%%%%%%%%%%%%%%%%
%Figure 4
%%%%%%%%%%%%%%%%%%%%%%%%%%%%%%%%%%%%%%%%%%%%%%%%%%%%%%%%%%%%%
\begin{figure}[t]
      \begin{center}
       \leavevmode
       \epsfxsize=8cm
       \epsfbox{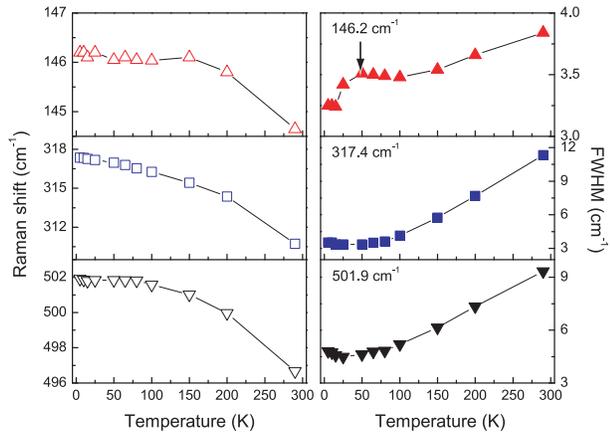}
        \caption{(Color online) (Left panel) Temperature dependence of
        the peak frequencies of selected phonons.
        (Right panel) Phonon linewidths as a function of temperature.} \label{fig.2}
\end{center}
\end{figure}
%%%%%%%%%%%%%%%%%%%%%%%%%%%%%%%%%%%%%%%%%%%%%%%%%%%%%%%%%%%%%

%%%%%%%%%%%%%%%%%%%%%%%%%%%%%%%%%%%%%%%%%%%%%%%%%%%%%%%%%%%%%
%Figure 5
%%%%%%%%%%%%%%%%%%%%%%%%%%%%%%%%%%%%%%%%%%%%%%%%%%%%%%%%%%%%%
\begin{figure}[t]
      \begin{center}
       \leavevmode
      \epsfxsize=7cm \epsfbox{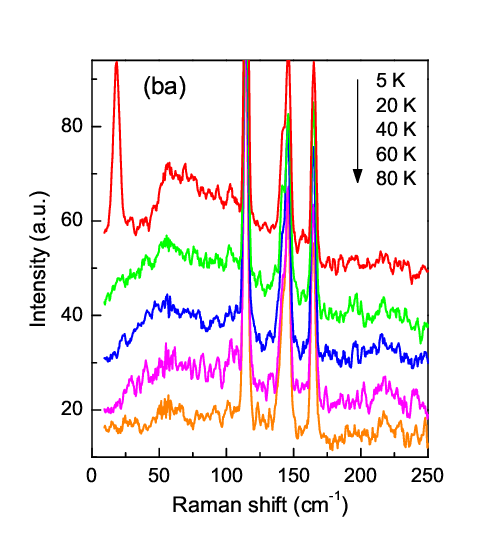}
        \caption{(Color online) Temperature dependence of magnetic
        Raman spectra in $ba$-polarization at 5, 20, 40, 60, and
        80 K (from upper to lower curve).} \label{fig.3}
\end{center}
\end{figure}
%%%%%%%%%%%%%%%%%%%%%%%%%%%%%%%%%%%%%%%%%%%%%%%%%%%%%%%%%%%%%%%%%%%%

%%%%%%%%%%%%%%%%%%%%%%%%%%%%%%%%%%%%%%%%%%%%%%%%%%%%%%%%%%%%%
%Figure 6
%%%%%%%%%%%%%%%%%%%%%%%%%%%%%%%%%%%%%%%%%%%%%%%%%%%%%%%%%%%%%
\begin{figure}[t]
      \begin{center}
       \leavevmode
       \epsfxsize=7cm \epsfbox{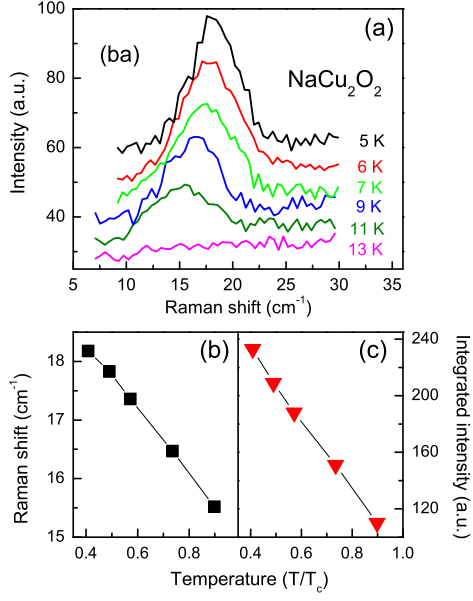}
        \caption{(Color online) (a) Low-frequency magnetic Raman spectra
        in $ba$-polarization as a function of temperature. (b,c)
        Temperature dependence of the peak position and integrated
        intensity on a reduced temperature scale, respectively.} \label{fig.4}
\end{center}
\end{figure}
%%%%%%%%%%%%%%%%%%%%%%%%%%%%%%%%%%%%%%%%%%%%%%%%%%%%%%%%%%%%%%%%%

%%%%%%%%%%%%%%%%%%%%%%%%%%%%%%%%%%%%%%%%%%%%%%%%%%%%%%%%%%%%%
%Figure 7
%%%%%%%%%%%%%%%%%%%%%%%%%%%%%%%%%%%%%%%%%%%%%%%%%%%%%%%%%%%%%
\begin{figure}[t]
      \begin{center}
       \leavevmode
       \epsfxsize=8cm
       \epsfbox{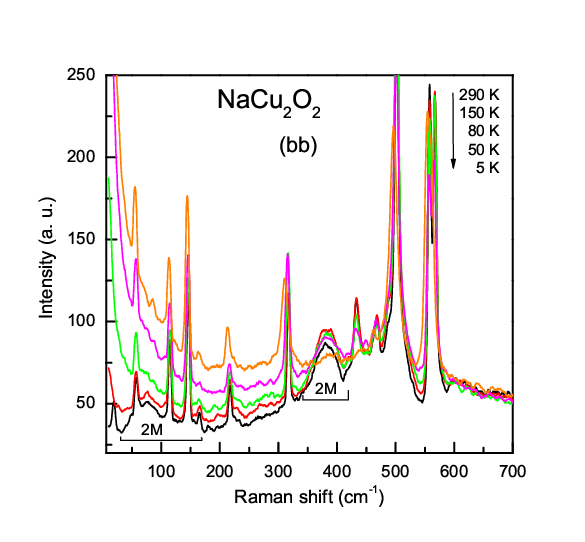}
        \caption{(Color online) Raman spectra
        in $bb$-polarization as a function of temperature;
        290, 150, 80, 50, and 5 K from upper to lower curve.} \label{fig.5}
\end{center}
\end{figure}
%%%%%%%%%%%%%%%%%%%%%%%%%%%%%%%%%%%%%%%%%%%%%%%%%%%%%%%%%%%%%%%

%%%%%%%%%%%%%%%%%%%%%%%%%%%%%%%%%%%%%%%%%%%%%%%%%%%%%%%%%%%%%
%Figure 8
%%%%%%%%%%%%%%%%%%%%%%%%%%%%%%%%%%%%%%%%%%%%%%%%%%%%%%%%%%%%%
\begin{figure}[tb]
      \begin{center}
       \leavevmode
       \epsfxsize=8cm
       \epsfbox{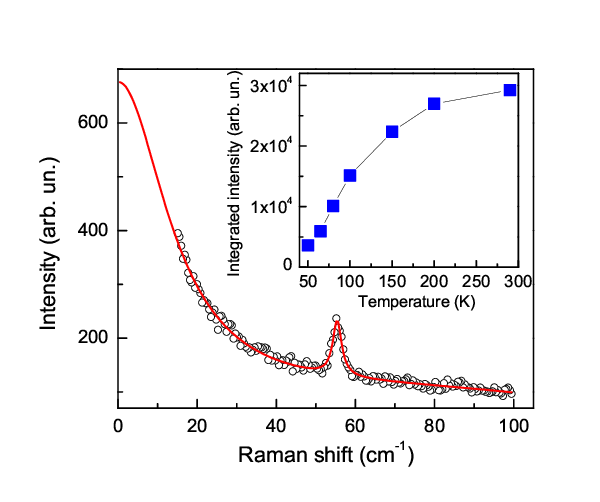}
        \caption{(Color online) Result of a fit of the quasielastic
response to a Lorentzian profile at 295 K. Inset: Temperature
dependence of the integrated scattering intensity of this
profile.} \label{fig.5}
\end{center}
\end{figure}

%%%%%%%%%%%%%%%%%%%%%%%%%%%%%%%%%%%%%%%%%%%%%%%%%%%%%%%%%%%%%%%%

\end{document}